\documentclass{revtex4}

\def\giorno{10/12/2017}

\usepackage{color}
\usepackage{amsmath,amsfonts,amssymb}

\def\a{\alpha}
\def\b{\beta}
\def\ga{\gamma}
\def\de{\delta}   %% NON ridefinire come \d !!!!
\def\eps{\varepsilon}
\def\vphi{\varphi}

\def\s{\sigma}

\def\vphi{\varphi}

\def\R{{\bf R}}

\def\De{\Delta}

\def\pa{\partial}

\def\o+{\oplus}

\def\ss{\subset}

\def\<{\langle}
\def\>{\rangle}

\def\({\left(}
\def\){\right)}
\def\[{\left[}
\def\]{\right]}
\def\=#1{\bar #1}
\def\~#1{\widetilde #1}
\def\wt#1{\widetilde #1}
\def\.#1{\dot #1}
\def\^#1{\widehat #1}
\def\wh#1{\widehat #1}
\def\"#1{\ddot #1}

\def\interno{\hskip 2pt \vbox{\hbox{\vbox to .18
truecm{\vfill\hbox to .25 truecm
{\hfill\hfill}\vfill}\vrule}\hrule}\hskip 2 pt}

\def\eeq{\end{equation}}
\def\beq{\begin{equation}}

\def\beql#1{\begin{equation} \label{#1}}

\def\eqref#1{(\ref{#1})}

\def\EOR{ \hfill $\odot$ \medskip}

\def\EOP{ \hfill $\triangle$ \medskip}

\def\g{\mathfrak{g}}

\def\symmref{AVL,CGbook,Olver1,Olver2,Stephani}
\def\SDES{Evans,Freedman,Ikeda,Kampen,Oksendal}

\begin{document}

\title{Symmetry and integrability for stochastic differential equations}

\author{G. Gaeta\thanks{ORCID 0000-0003-3310-3455; e-mail: {\tt giuseppe.gaeta@unimi.it}} \ \ and C. Lunini\thanks{ORCID 0000-0002-2643-1226; e-mail: {\tt claudia.lunini@studenti.unimi.it}} \\
{\it Dipartimento di Matematica, Universit\`a degli Studi di
Milano} \\ {\it via Saldini 50, 20133 Milano (Italy)} \\
{\tt \giorno } }

%\date{{\tt split version} -- \giorno }

\begin{abstract}
We discuss the interrelations between symmetry of an Ito
stochastic differential equations (or systems thereof) and its
integrability, extending in party results by R. Kozlov [{\it J.
Phys. A} {\bf 43} (2010) \& {\bf 44} (2011)].  Together with integrability,
we also consider the relations between symmetries and reducibility
of a system of SDEs to a lower dimensional one. We consider both
``deterministic'' symmetries and ``random'' ones, in the sense
introduced recently by Gaeta and Spadaro [{\it J.Math. Phys.} {\bf
58} (2017)].
\end{abstract}

\maketitle

\section{Introduction}

It is well known that symmetry methods are among the most powerful
tools for the study of nonlinear deterministic differential
equations \cite{\symmref}. It is thus natural to think they could
be useful also in the study of \emph{stochastic}  differential
equations (SDEs).

This observation is of course not new, and indeed there is by now
a substantial literature devoted to symmetries of SDEs (see
\cite{GGPR} for an extended  reference list).  In a first phase
the efforts focused on determining the proper -- that is, useful
-- definition of symmetry for a SDE and hence the
\emph{determining equations}; there is now a consensus on what
this suitable definition is (see below). The second and crucial
phase is of course to understand how these can be \emph{used} in
the study of SDEs.

The main idea -- paralleling the approach for deterministic
equations -- is to use symmetry-adapted coordinates to implement a
\emph{reduction} of the SDE. It should be stressed that here one
should think of the decomposition of a given system into a smaller
(lower dimensional) one plus one or more \emph{reconstruction
equations}, as in the case of deterministic Dynamical Systems;
we are here going to study systems of Ito equations, which are the stochastic
counterpart of first order ODEs, i.e. indeed Dynamical Systems.

A key result in this direction was obtained by Kozlov \cite{Koz1}
for scalar SDEs, showing that if such an SDE possesses a symmetry
of a certain type (``simple'' symmetries, to be defined below),
then it can be integrated; the theorem is actually constructive,
i.e. the symmetry determines a change of variables which allows
for explicit integration of the SDE (see below for details).

In the case of systems of SDE, Kozlov's approach \cite{Koz2,Koz3} shows
that a symmetry of the appropriate type implies that the system
can be ``partially integrated'' (in a sense to be made precise
below).

Kozlov approach is based -- like analogous results for
deterministic equations -- on changes of variables to use
favorable properties of symmetry-adapted coordinates; it is
essential that symmetries are preserved under changes of
coordinates. In a companion paper \cite{GLjnmp} we have argued
that while this fact is entirely obvious for deterministic
equations, preservation of symmetries for Ito stochastic equations
cannot be given for granted. The reason is that Ito equations are
not geometrical objects, and indeed do not transform in the
``usual'' way (i.e. under the familiar chain rule) under changes
of coordinates: in fact, they transform according to the Ito rule.
However it turns out that there is a rather ample class of
symmetries which \emph{are} preserved under changes of
coordinates, and these include the kind of symmetries relevant for
Kozlov's theory; the reader is referred to our work \cite{GLjnmp}
for a discussion, while the results relevant to our present
subject are  recalled -- after discussing in Section \ref{sec:symmsde}
some general features of symmetries of SDEs -- in Section
\ref{sec:ItoStrat}.

Equipped with these preliminary results, we pass to study the
relation between symmetry and integration -- or at least reduction
-- for (Ito) SDEs. For what concerns a single symmetry, the
canonical result in the literature, already recalled above, is due
to Kozlov \cite{Koz1} and says that the existence of a symmetry of
a given type is a \emph{sufficient} condition for the
integrability of an Ito scalar SDE. We will show that the
condition is not only sufficient but also \emph{necessary}, see Section
\ref{sec:Kozscalar}.

We will then pass to consider the case of multi-dimensional
systems of SDEs with a single symmetry and discuss the reduction
which is possible in this case (see Section \ref{sec:Kozsystems});
this will also give an idea of the kind of results which can be
obtained in the general case.

Finally (in Section \ref{sec:multiple}) we will consider the case
of a $n$-dimensional system of Ito SDEs with a $q$-dimensional
solvable symmetry algebra (the restriction to considering
\emph{solvable} symmetry algebras descends from well known results
in the study of deterministic equations and their symmetry
reduction \cite{\symmref}), and discuss how the symmetry allows
for a multiple-stages reduction of the system; in general this is
actually setting the system in a ``partially triangular'' form,
with an $m$-dimensional (with $m <n$) system and a number ($q =
n-m$) of ``reconstruction equations''. Here again we will compare
our results with those obtained by Kozlov \cite{Koz2,Koz3}.

So far we have considered maps acting by transformation of the
dependent variables, possibly depending on time; we will also show
(in Section \ref{sec:rankoz}, with examples in Section \ref{sec:exarandom}) that the results discussed so far
can be extended to the case of \emph{random maps} \cite{Arnbook,ArnImk,GS}, i.e.
transformations depending on the Wiener processes entering in the
equations under study. In this case the transformation is subject
to certain restrictions to guarantee we remain within the category
of Ito equations, and a number of other details should be
controlled. For this setting, we will consider only \emph{scalar}
equations, postponing the study of systems to a future
contribution. It will turn out that it is possible to obtain
a nearly full extension -- at least for scalar equations -- of
Kozlov theory to the setting of random maps and random symmetries; the only difference being that now an additional compatibility condition \eqref{eq:bcomp} involving the coefficients of the Ito equation and of the symmetry vector field should be satisfied.

We also provide two Appendices; the first shows that albeit one could apparently extend Kozlov's theory to other types of integrable SDEs (separable ones), this framework is already contained in the ``standard'' case; in the second we discuss when one has infinitely many (simple) random symmetries for a given scalar Ito equation.

Finally, it should be mentioned that the literature also considers
(systems of) higher order SDEs and their symmetries
\cite{GGPR,SMS}; we will not consider these, neither other
possible generalizations of the setting (i.e., that of systems of
Ito equations) we have chosen to study.
Similarly, when dealing with deterministic equations one also
considers more general classes of symmetries than Lie-point ones
\cite{\symmref}; but we will not consider these here.

\section{Symmetry of stochastic differential equations}
\label{sec:symmsde}

In this note, by ``differential equation'' (DEs) we will always
mean possibly a vector one, i.e. a system of differential
equations, unless the contrary is explicitly stated (in this case
we speak of a scalar equation).

Given a deterministic DE $\Delta$ of order $n$ in the basis
manifold $M$ (this include dependent and independent variables),
it is well known that this is identified with its \emph{solution
manifold} $S_\Delta$ in the Jet bundle $J^n M$. A Lie-point
symmetry (generator) is then a vector field $X$ on $M$ which, once
\emph{prolonged} to the vector field $X^{(n)}$ on $J^n M$, is
tangent to $S_\De \ss J^n M$ \cite{\symmref}.

Both submanifolds and vector fields are geometrical object; they
are defined independently of any coordinate system, and it is thus
entirely obvious that tangency relations -- hence in particular
$X$ being a symmetry for $\Delta$ -- hold independently of
coordinates. This also means that symmetries are preserved under
changes of coordinates.

Let us consider an (Ito) stochastic differential equation (SDE)
\cite{Arnbook,\SDES} \beql{eq:Ito} d x^i \ = \ f^i (x,t) \, d t \
+ \ \s^i_{\ j} (x,t) \, d w^j \ ; \eeq in \eqref{eq:Ito} and
below, $t \in R$, $x \in R^n$ (the $x^i$ can be thought as local
coordinates on a smooth manifold, as we take care of
distinguishing covariant and contravariant indices albeit never
introducing explicitly a metric), $f$ and $\s$ are smooth vector
and matrix functions of their arguments, and $w^j$ ($j = 1,...,m$)
are independent standard Wiener processes.

For Ito equations, a \emph{geometrical} interpretation of the
equation is missing, and one is forced to resort to a purely
\emph{algebraic} notion of symmetry. Actually, not only a
geometrical interpretation is missing, but this is not possible:
in fact Ito equations and more generally Ito differentials do not
transform in the ``usual'' way, i.e. according to the familiar
chain rule, under changes of coordinates; they transform instead
according to the Ito rule, and this is inherent to their very
nature \cite{Arnbook,\SDES,Stroock} (see \cite{Schulman}, chapter
5, for a Physics point of view).

\medskip\noindent
{\bf Remark 1.} The situation is different for Stratonovich SDEs;
actually the main motivation for their introduction was precisely
to have SDEs transforming ``nicely'' under changes of coordinates.
For these a geometrical description is indeed possible, and hence
symmetries are trivially preserved under any change of coordinates
(see also the detailed discussion in \cite{GLjnmp}). On the other
hand, Stratonovich equations present several problems from the
probabilistic point of view; see e.g.
\cite{Arnbook,\SDES,Stroock}. It should be stressed that albeit
there is a correspondence between Ito and Stratonovich equations
(which is not entirely trivial, see e.g. the discussion in
\cite{Stroock}), there is not an identity between general
symmetries -- even of deterministic type \cite{Unal} -- of an Ito
equation and those of the corresponding Stratonovich one. It is
instead known that such an identity holds for (both deterministic
\cite{Unal} and random \cite{GLjnmp}) \emph{simple} symmetries
(see again \cite{GLjnmp} for discussion and examples). Note that
previous work by one of the present authors \cite{GGPR,GS}
contains wrong statements in this respect (originating in a
trivial mistake in the definition of the Ito Laplacian, as
discussed in \cite{GLjnmp}). \EOR

\medskip\noindent
{\bf Remark 2.} It should be stressed  that, in view of the
non-covariance of Ito equations, and of the purely algebraic
definition of symmetries for them (see e.g. the review \cite{GGPR}
for details) it is not at all guaranteed {\it a priori} that
symmetries will be preserved under changes of coordinates. An
analysis of this problem was provided recently \cite{GLjnmp}, and
it turned out that the symmetries which are relevant for
integrability and/or reduction are preserved. Such results
will be recalled below, and are based on the relation with symmetries of the corresponding Stratonovich equation, see Remark 1. \EOR
\bigskip

One could, in principles, consider general mappings and vector
fields in the $(x,t,w)$ space -- albeit with some restrictions if
these have to make physical sense, see e.g. \cite{GGPR,GS}. That
is, we could generally speaking consider maps
$$ (x,t;w) \ \to \ \( \wt{x} (x,t,w) , \wt{t} (x,t,w) ; \wt{w}
(x,t,w) \) \ . $$

Within these, we will denote by {\bf simple} the maps (and vector
fields) which act only on the spatial variables, leaving the time
$t$ (and the Wiener processes) unchanged. We will also denote as
{\bf deterministic} the maps (and vector fields) which only depend
on $x$ (and possibly $t$), as opposed to the {\bf random} ones (to
be considered in subsection \ref{sec:simpleran} below), in which
dependence on the Wiener processes $w$ is also present.

Within our present context, we are interested only in
\emph{simple} maps. The reason for this is that only these are
relevant in Kozlov theory, as discussed in his works
\cite{Koz1,Koz2,Koz3} and as will be recalled below. This simplifies considerably our task.

\subsection{Simple deterministic maps \& symmetries}
\label{sec:simpledet}

Let us first consider a (simple) smooth\footnote{By ``smooth'' we
will always mean $C^\infty$, albeit in most steps it would be
sufficient to consider $C^2$ smoothness.} map \beql{eq:smap} (x,t)
\ \to (\wt{x} , t ) \ , \ \ \wt{x} = \wt{x} (x,t) \ ; \eeq this
induces a map $d x \to d \wt{x}$ and hence a map on SDEs
\eqref{eq:Ito}; in particular if \eqref{eq:smap} is (locally)
inverted to give \beq x \ = \ \Phi (\wt{x},t ) \ , \eeq then by
Ito formula \beq d x^i \ = \ \( \frac{\pa \Phi^i}{\pa \wt{x}^j} \)
\, d \wt{x}^j \ + \ \( \frac{\pa \Phi}{\pa t} \) \, d t \ + \
\frac12 \( \frac{\pa^2 \phi^i}{\pa \wt{x}^j \pa \wt{x}^k} \)
\wt{\s}^j_{\ \ell} \wt{\s}^k_{\ \ell} \, d t \ ; \eeq similarly
the functions $f^i (x,t)$ and $\s^i_{\ j} (x,t)$ are mapped into
functions $\wt{f}^i (\wt{x},t ) $ and $\wt{\s}^i_{\ j} (\wt{x} , t
)$. In this way, \eqref{eq:Ito} is mapped into a new Ito equation
\beql{eq:Ito2} d \wt{x}^i \ = \ \wh{f}^i
(\wt{x},t) \, d t \ + \ \wh{\s}^i_{\ j} (\wt{x},t) \, d w^j \ ;
\eeq note that here the
$\wh{f}$ and $\wh{\s}$ take into account not only the change of
their variables, but also the contribution arising from $d x$
expressed in the new variables via the Ito formula.

We say that \eqref{eq:smap} is a {\bf symmetry} for \eqref{eq:Ito}
if \eqref{eq:Ito2} is identical to \eqref{eq:Ito}, i.e. if the ($n
+ m \cdot n)$ conditions \beq \wh{f}^i (x,t) \ = \ f^i (x,t) \ , \
\ \wh{\s}^i_{\ j} (x,t) \ = \ \s^i_{\ j} (x,t) \eeq are satisfied
identically in $(x,t)$.

We are specially interested in the case where the map
\eqref{eq:smap} is a near-identity one, $\wt{x}^i = x^i + \eps
(\de x)^i$ and can hence be seen as the infinitesimal action of a
vector field $X$, \beql{eq:X} X \ = \ \vphi^i (x,t) \,
\frac{\pa}{\pa x^i} \ \equiv \ \vphi^i (x,t) \, \pa_i \ . \eeq
(Note that $X$ has no component along the $t$ variable; this
corresponds to the restriction made above, see eq.\eqref{eq:smap},
for simple symmetries.) In this case we speak of a
\emph{Lie-point} (simple) symmetry.

We obtain then easily (the reader is referred to \cite{GRQ1} for
details and explicit computations) the \emph{determining equations
for (simple, deterministic) Lie-point symmetries of SDEs}:
\begin{eqnarray}
\vphi^i_t \ + \ f^j \, (\pa_j \vphi^i) \ - \ \vphi^j \, (\pa_j f^i) &=&
- \, \frac12 \ (\Delta \vphi^i ) \ , \label{eq:deteqIto1}  \\
\s^j_{\ k} \, (\pa_j \vphi^i) \ - \ \vphi^j \, (\pa_j \s^i_{\ k} )
&=& 0 \ . \label{eq:deteqIto2} \end{eqnarray} Here and below
$\Delta$ denotes the \emph{Ito Laplacian}, which in general (for
functions possibly depending also on the $w^k$, see next
subsection) is defined as \beql{eq:Delta} \Delta f \ = \
\delta_{ik} \ \[ \( \frac{\pa^2 f}{\pa w^i \pa w^k} \) \ + \
\s^j_{\ i} \s^m_{\ k} \, \( \frac{\pa^2 f}{\pa x^j \pa x^m} \) \ +
\ 2 \, \s^j_{\ i} \, \( \frac{\pa^2 f}{\pa x^j \pa w^k} \) \] \ .
\eeq

\subsection{Simple random symmetries}
\label{sec:simpleran}

We can consider more general transformations; in particular one
can consider maps and vector fields which depend on the Wiener
processes realizations $w^k$. This approach (actually in a more
general setting) was first considered by Arnold and Imkeller
\cite{Arnbook,ArnImk} in the context of \emph{normal forms} for
SDEs; random symmetries of SDEs are considered e.g. in \cite{GS}
(see also \cite{GGPR}), to which we refer for explicit
computations and details.

For what concerns our discussion here, we consider \emph{simple
random maps}, \beql{eq:srmap} \( x,t;w \) \ \to \( \wt{x} , t ; w
\) \ , \ \ \wt{x} = \wt{x} (x,t;w) \ ; \eeq and the corresponding
symmetries, i.e. \emph{simple random symmetries}. These are the
vector fields \beql{eq:Xran} X \ = \ \vphi^i (x,t;w) \, \pa_i \eeq
leaving the equation \eqref{eq:Ito} invariant

It is shown in \cite{GS} that, using the notation, to be used
routinely in the following, \beql{eq:pahat} \^\pa_k \ := \
\frac{\pa}{\pa w^k} \ , \eeq the \emph{determining equations for
simple random Lie-point symmetries of a SDE} \eqref{eq:Ito} read
\begin{eqnarray}
(\pa_t \vphi^i) \ + \ f^j \, (\pa_j \vphi^i) \ - \ \vphi^j \, (\pa_j f^i) &=&
- \, \frac12 \ (\Delta \vphi^i ) \ , \label{eq:deteqItoR1} \\
(\^\pa_k \vphi^i) \ + \ \s^j_{\ k} \, (\pa_j \vphi^i) \ - \
\vphi^j \, (\pa_j \s^i_{\ k} ) &=& 0 \ . \label{eq:deteqItoR2}
\end{eqnarray}

\section{Preservation of simple symmetries of Ito equations}
\label{sec:ItoStrat}

Ito equations are not geometrical objects. In fact, under changes
of coordinates they do not transform in the ``usual'' way, i.e.
under the chain rule, but in their own way, i.e. under the Ito
rule.

This means that symmetries are defined in an \emph{algebraic}
rather than \emph{geometrical} way; and, at difference with
deterministic differential equations, it is not obvious that if we
consider an Ito equation $E$, a vector field $X$ which is a
symmetry for $E$, and a change of coordinates (mapping the
equation $E$ into an equation $\wt{E}$), the vector field $X$ will
also be a symmetry for the transformed equation $\wt{E}$. This
point was raised and discussed in a recent paper of ours
\cite{GLjnmp}.

The situation is however quite different -- and similar to the
familiar one for deterministic equations -- if one considers, as
in Kozlov theory, only \emph{simple symmetries}.

\medskip\noindent
{\bf Lemma 1.} {\it Simple (random or deterministic) symmetries of
an Ito equation are preserved under changes of coordinates.}

\medskip\noindent
{\bf Proof.} See \cite{GLjnmp}.
\EOP

%\newpage

\section{Deterministic symmetry and integrability for a scalar Ito SDE}
\label{sec:Kozscalar}

%\subsection{The two-ways Kozlov Theorem}
%\label{sec:twoways}

We will consider the following result, which in its essential part is due to Kozlov:

\medskip\noindent
{\bf Theorem 1.} {\it The SDE \beql{eq:koz2} d y \ = \ \wt{f}
(y,t) \ d t \ + \ \wt{\s} (y,t) \ d w  \eeq can be transformed by
a deterministic map into \beql{eq:koz1} d x \ = \ f(t) \, d t \ +
\ \s (t) \, d w \ , \eeq and hence explicitly integrated, \emph{if
and only if} it admits a simple deterministic symmetry.

If the generator of the latter is \beql{eq:symm0} X \ = \ \vphi
(y,t) \ \pa_y \ , \eeq then the change of variables $y  =  F
(x,t)$ transforming \eqref{eq:koz2} into \eqref{eq:koz1} is the
inverse to the map $x = \Phi (y,t)$ identified by
\beql{eq:covrel2} \Phi (y,t) \ = \ \int \frac{1}{\vphi (y,t) } \ d
y \ . \eeq}

\medskip\noindent
{\bf Proof.} Consider a (scalar) SDE of the form \eqref{eq:koz1}.
This is obviously and elementarily integrable; moreover it admits
the Lie symmetry generator \beql{eq:symm} X \ = \ \pa_x \ . \eeq

If we operate  a change of variable of the form \beql{eq:cov} x \
= \ \Phi (y,t) \eeq (note we are leaving $t$ and $w=w(t)$
untouched), the Ito formula gives \beq d x \ = \ (\pa_t \Phi) \, d
t \ + \ (\pa_y \Phi) \, d y \ + \ \frac12 \, (\pa_y \Phi )^2 \,
\s^2 (t) \, d t \ . \eeq Putting together this and \eqref{eq:koz1}
we get \beq \Phi_y \, d y \ = \ \[ f(t) \ - \ \Phi_t \ - \ \frac12
\, \Phi_{yy} \, \wt{\s}^2 (t) \] \, d t \ + \ \s(t) \, d w \ .
\eeq Now we just note that $\Phi_y \not= 0$, or \eqref{eq:cov}
would not be a proper change of variable, so that we can divide by
$\Phi_y$. Thus \eqref{eq:koz1} reads, in terms of the new
variable, precisely as \eqref{eq:koz2}, where we have written
\begin{eqnarray} \wt{f} (y,t) & := &  \frac{1}{\Phi_y} \ \[ f \ - \ \Phi_t
\ - \ \frac12 \, \Phi_y^2 \, \s^2  \] \ ; \label{eq:koz2F} \\
\wt{\s} &:=& \frac{\s}{\Phi_y} \ . \label{eq:koz2S}
 \end{eqnarray}

On the other hand, in the new coordinates the vector field $X$
reads \beql{eq:symm2} X \ = \ \frac{\pa y}{\pa x} \ \pa_y \ = \
\frac{\pa F (x,t)}{\pa x} \, \frac{\pa}{\pa y} \ = \
\frac{1}{\Phi_y} \ \pa_y \ . \eeq

Following the discussion in Sect.\ref{sec:ItoStrat}, we are
guaranteed by Lemma 1 that \eqref{eq:symm2} is a symmetry for
\eqref{eq:koz2}.

In view of the general nature of \eqref{eq:cov}, we have thus
shown that if the equation \eqref{eq:koz2} can be mapped via a
simple change of variables of the form \beql{eq:cov2} y \ = \ F
(x,t) \ , \eeq to the equation \eqref{eq:koz1}, then necessarily
it enjoys the symmetry \eqref{eq:symm2}. The required change of
variables \eqref{eq:cov2} is just the inverse to \eqref{eq:cov},
i.e. these satisfy \beql{eq:covrel} \Phi [ F (x,t),t] = x \ ; \ \
F [ \Phi (y,t), t] =  y \ . \eeq

It just remains to identify $F$ in terms of the coefficient $\vphi
(y,t) $ in \eqref{eq:symm0}. Comparing the latter and
\eqref{eq:symm2} we immediately have \beql{eq:phiPhi} \vphi (y,t)
\ = \ \frac{1}{\Phi_y (y,t)} \ , \eeq which shows that
\eqref{eq:covrel2} holds and identifies $\Phi (y,t)$ in terms of
$\vphi (y,t)$. As we need the change of variables leading from $y$
to $x$, we just need the change of variable $y = F(x,t)$ inverse
to $\Phi$, as stated in \eqref{eq:covrel}. This completes the
proof in one direction, i.e. shows that indeed symmetry is a
necessary condition for the equation \eqref{eq:koz2} to be
integrable.

In order to show that the condition is also sufficient for the
integrability of \eqref{eq:koz2}, it suffices to perform the
change of variables, thus obtaining \eqref{eq:koz1} which is
obviously integrable. \EOP

\medskip\noindent
{\bf Remark 3.} The Theorem is due to Kozlov \cite{Koz1}, who
stated it in one direction only -- i.e. identifying the symmetry
condition as a \emph{sufficient} one to guarantee integrability --
and gave it in constructive form, i.e. identifying the change of
variables which makes the equation explicitly integrable. Our
contribution here is to show that this condition is not only
sufficient but also \emph{necessary}. \EOR

\medskip\noindent
{\bf Remark 4.} One could think of extending the Kozlov theorem to
more general situations: in fact, stochastic equations of the form
\eqref{eq:koz1} are not the only ones which can be integrated, and
one could e.g. consider also those of the form $ d x = \b (x) \, [
f(t) \ d t \ + \ \s (t) \ d w ]$. This can be done, but actually
gives no new result. The discussion of this fact is given in
\ref{sec:extended}. \EOR

\medskip\noindent
{\bf Remark 5.} The most relevant feature of Kozlov theorem is of
course that it is \emph{constructive}: once we have determined the
simple symmetry (if any) of the equation, it suffices to invert
\eqref{eq:phiPhi}, which of course yields \eqref{eq:covrel2}, to
explicitly get the required change of variables and explicitly
integrate the equation. \EOR

\medskip\noindent
{\bf Remark 6.} The function $\Phi (y,t)$ is determined by
\eqref{eq:covrel2} up to an ``integration constant'' which is
actually an arbitrary function of $t$; it is clear from
\eqref{eq:koz2F}, \eqref{eq:koz2S} that the only effect of this
would be to add a function of $t$ alone to $F(y,t)$ -- or, seen
from the other end of the procedure, to $f(t)$ -- while leaving
$S(y,t)$ and $\s (t)$ unchanged; thus with no loss of generality
for what concerns the integration of the SDE under study. (Similar
considerations would also apply for the results in the next
section \ref{sec:Kozsystems}, and will not be repeated there.)
\EOR

\medskip\noindent
{\bf Example 1.} The Ito equation\footnote{This and the following
example are also given in \cite{GLjnmp}; see there for details.}
\beql{eq:example1} d y \ = \ \[ e^{- y} \ - \ (1/2) \, e^{-2 y} \]
\, d t \ + \ e^{- y} \, d w \eeq admits the vector field $ X  =
e^{- y} \pa_y $ as a (Lie-point) symmetry generator. By the change
of variables $x = \exp[y]$ the vector field reads $X = \pa_x$, and
the initial equation \eqref{eq:example1} reads
$$ d x \ = \ d t \ + \ d w \ . \eqno{\odot} $$

\medskip\noindent
{\bf Example 2.} The (quite involved) equation \beql{eq:example2}
d y \ = \ \frac{e^{-t} \, (1 + y^2)^2}{8 y^3} \, \( - 4 y^2 \, +
\, e^t (3 y^4 +2 y^2 - 1) \) \, d t \ - \ \frac{(1+y^2)^2}{2 y} \,
d w \eeq admits the vector field $ X = - [ (1+y^2)^2/(2 y)] \pa_y$
as a symmetry. Correspondingly, passing to the variable $ x =1/(1+
y^2)$, we get $X = \pa_x$ and the equation \eqref{eq:example2}
just reads
$$ d x \ = \ e^{- t} \, d t \ + \ d w \ . \eqno{\odot} $$

\section{Deterministic symmetry and reduction for a system of Ito SDEs}
\label{sec:Kozsystems}

In the previous section we have considered reduction of a scalar
Ito equation under a simple Lie-point symmetry. In view of the
discussion given there it is not surprising that the same result
-- with some obvious changes -- also holds when we consider a
system of Ito SDEs with a simple symmetry. In this section we will
discuss in detail such an extension. We will again show that
Kozlov's sufficient conditions are also necessary.

We thus consider a system of $n$ SDEs for $x^1 (t),...,x^n (t)$,
which enjoys a symmetry of an appropriate type, and want to reduce
this to a system of $n-1$ equations for $y^1 (t), ... , y^{n-1}
(t)$; the processes $y^i (t)$ will be defined in terms of the $x^i
(t)$. We have the following result.

\medskip\noindent
{\bf Theorem 2.} {\it Consider the system \eqref{eq:Ito}, with
$i=1,...,n$. The \emph{necessary and sufficient} condition for the
existence of a simple deterministic non-degenerate change of
variables \eqref{eq:smap} which brings the system into the form
\beq \label{eq:reduced} d{y}^i \ = \ g^i ({y}^1, . .
.,{y}^{n-1},t) \, d t \ + \   \rho^i_{\ k} (y^1,...,y^{n-1},t) \,
d w^k(t) \ \ \ \ (i=1,...,n ) \eeq is that \eqref{eq:Ito} admits a
simple deterministic Lie-point symmetry \eqref{eq:X}.}

\medskip\noindent
{\bf Proof.} We will proceed substantially in the same way as in
the proof for Theorem 1. We start from the ``reduced'' system
\eqref{eq:reduced}, which is clearly invariant under the action of
the vector field $X = (\pa / \pa y^n)$; we consider a general
(obviously, invertible) change of variables, i.e. we bring it to
its general form in arbitrary coordinates. So we consider the
change of coordinates $(y^1 , ... , y^n) \to (x^1 , ... , x^n)$
defined by \beql{eq:Phi} y^i \ = \ F^i (x,t) \ , \eeq with inverse
\beql{eq:FF} x^i \ = \ \Phi^i (y,t) \ . \eeq Under this change of
variables the system \eqref{eq:reduced} becomes (using Ito rule)
\beql{eq:reda1} dy^i \ = \ d F^i(x,t) \ = \ \frac{\pa F^i}{\pa
x^k} \, dx^k \ + \  \frac{\pa F^i}{\pa t} \, dt \ + \ \frac{1}{2}
\, \frac{\pa^2 F^i}{\pa x^k\pa  x^\ell} \, dx^k \, dx^\ell \ .
\eeq Let us focus on the last term, and more specifically on the
factor $d x^k d x^\ell$; using \eqref{eq:FF} and again Ito rule,
\begin{eqnarray*} d x^k \, d x^\ell &=& \frac{\pa \Phi^k}{\pa y^m} \,
\frac{\pa \Phi^\ell}{\pa y^s} \ d y^m \, d y^s \ + \ o (d t) \\
&=& \frac{\pa \Phi^k}{\pa y^m} \, \frac{\pa \Phi^\ell}{\pa y^s} \
\rho^m_{\ p} \, \rho^s_{\ q} \, d w^p \, d w^q \ + \ o (dt) \\
&=& \frac{\pa \Phi^k}{\pa y^m} \, \frac{\pa \Phi^\ell}{\pa y^s} \
\rho^m_{\ q} \, \rho^s_{\ q} \, d t \ + \ o (dt) \ . \end{eqnarray*}

Inserting this into \eqref{eq:reda1}, the latter reads
\beql{eq:reda2}
dy^i \ = \ \frac{\pa F^i}{\pa x^k} \, dx^k \ + \  \frac{\pa F^i}{\pa t} \, dt \ + \ \frac{1}{2} \, \frac{\pa^2 F^i}{\pa x^k\pa  x^\ell} \,  \( \frac{\pa \Phi^k}{\pa y^m} \, \frac{\pa \Phi^\ell}{\pa y^s} \ \rho^m_{\ q} \, \rho^s_{\ q} \) \, d t\ . \eeq

These allow to deduce the SDEs satisfied by the process in terms of the new (i.e. the $x$) variables; in fact, using also \eqref{eq:reduced}, the above relation \eqref{eq:reda2} is rewritten as
\beq \label{eq:jac1}
\frac{\pa F^i}{\pa x^k} \, dx^k \ = \ \[ g^i \, - \, \frac{\pa F^i}{\pa t} \, - \, \frac{1}{2} \, \frac{\pa^2 F^i}{\pa x^k\pa  x^l} \, \( \frac{\pa \Phi^k}{\pa y^m} \, \frac{\pa \Phi^\ell}{\pa y^s} \ \rho^m_{\ q} \, \rho^s_{\ q} \) \] \, dt \ + \ \rho^i_{\ k} \, dw^k(t) \ .
\eeq

In order to obtain a system of the form \eqref{eq:Ito}, we need to invert the matrix $(\pa F^i / \pa x^k)$, that is the Jacobian matrix of $F$. This inverse is of course the Jacobian for the inverse change of coordinates, i.e. the Jacobian for $\Phi$; in fact \beq \label{eq:delta}
\delta^i_k \ =\ \frac{\pa y^i}{\pa y^k} \ = \ \frac{\pa F^i[\Phi(y,t),t]}{\pa x^k} \ = \ \frac{\pa F^i}{\pa x^l} \, \frac{\pa \Phi^l}{\pa y^k} \ . \eeq

By acting on \eqref{eq:jac1} from the left with the matrix $(\pa F / \pa y) = (\pa F / \pa x)^{-1}$, we get
\begin{eqnarray}
    dx^k &=& \frac{\pa \Phi^k}{\pa y^i} \ \[ g^i \, - \, \frac{\pa F^i}{\pa t} \, - \, \frac{1}{2} \, \frac{\pa^2 F^i}{\pa x^m\pa  x^l} \, \( \frac{\pa \Phi^k}{\pa y^m} \, \frac{\pa \Phi^\ell}{\pa y^s} \ \rho^m_{\ q} \, \rho^s_{\ q} \) \] \, dt \ + \ \frac{\pa \Phi^k}{\pa y^i} \, \rho^i_\ell \, dw^\ell \nonumber \\        & := & f^k(x,t) \, dt \ + \ \s^k_{\ \ell} (x,t) \, dw^\ell \ . \label{eq:newsystem}
\end{eqnarray}
In the last step we have simply defined
\begin{eqnarray}
f^i (x,t) & := & g^i \, - \, \frac{\pa F^i}{\pa t} \, - \, \frac{1}{2} \, \frac{\pa^2 F^i}{\pa x^m\pa  x^l} \, \( \frac{\pa \Phi^k}{\pa y^m} \, \frac{\pa \Phi^\ell}{\pa y^s} \ \rho^m_{\ q} \, \rho^s_{\ q} \) \ , \label{eq:fnew} \\
\s^i_{\ k} (x,t) & := & \frac{\pa \Phi^i}{\pa y^j} \, \rho^j_{\ k} \ ; \label{eq:snew} \end{eqnarray}
here all the quantities which are function of $y$ should be thought as function of $x$ through \eqref{eq:Phi}.

Let us now turn our attention to symmetries. The original system
\eqref{eq:reduced} is by construction invariant under $X = (\pa /
\pa y^n)$. In the $x$ variables this vector field is described by
\beq \label{eq:Xnew} X \ = \ \( \frac{\pa x^i}{\pa y^n} \) \
\frac{\pa}{\pa x^i} \ = \ \frac{\pa \Phi^i}{\pa y^n} \,
\frac{\pa}{\pa x^i} \ := \ \vphi^i (x,t) \, \pa_i \ . \eeq It
should be stressed that when considering (possibly time-dependent)
vector fields acting in $\R^n$, we are dealing with the $x^i$,
$y^i$ as coordinates in $\R^n$, not with the stochastic processes
$x^i (t)$, $y^i (t)$ defined by SDEs \eqref{eq:reduced},
\eqref{eq:Ito}; thus we do \emph{not} have to use the Ito formula.

As discussed in Section \ref{sec:ItoStrat}, it is not guaranteed
{\it apriori} that symmetries present in one system of coordinates
will still be present in another one; however as we deal with
simple symmetries we are guaranteed by Lemma 1 we still have $X$
-- now given as \eqref{eq:Xnew} -- as a symmetry.

We have thus determined the most general SDE \eqref{eq:Ito} which
can be obtained from \eqref{eq:reduced} by a change of variables;
these are identified by \eqref{eq:fnew}, \eqref{eq:snew}.
Reversing our point of view, we have shown that all the SDEs
\eqref{eq:Ito} which can be reduced to the form \eqref{eq:reduced}
by a change of variables are characterized by $f$, $\s$ as given
in \eqref{eq:fnew}, \eqref{eq:snew}. Moreover these admit by
construction the Lie-point symmetry \eqref{eq:Xnew}. This shows
that indeed the presence of such a symmetry is a necessary
condition for the equation \eqref{eq:Ito} to be reducible to the
form \eqref{eq:reduced}.

Moreover, our discussion shows that there is a simple relation
between the symmetry \eqref{eq:Xnew} and the change of variables
needed to take \eqref{eq:Ito} into the reduced form
\eqref{eq:reduced}; this is given by \beql{eq:phiFF} \varphi^i \ =
\ \frac{\pa \Phi^i}{\pa y^n} \ . \eeq

In order to show that the presence of such a symmetry
\eqref{eq:Xnew} for the equation \eqref{eq:Ito} characterized by
\eqref{eq:fnew}, \eqref{eq:snew} is also a sufficient condition
for its (reducibility to the form \eqref{eq:reduced} and hence)
integrability, it suffices to note (as in \cite{Koz1}) that the
change of variables \eqref{eq:Phi} with $\Phi$ given by \eqref{eq:phiFF}
produces exactly the required reduction. \EOP

\medskip\noindent
{\bf Remark 7.} If $( y^1 (t),...,y^{n-1} (t) )$ are a solution to
the autonomous subsystem consisting of the first $n-1$ equations
in \eqref{eq:reduced}, then $y^n (t)$ is given by
\begin{eqnarray}
y^n (t) &=& y^n(t_0) \ + \ \int_{t_0}^t g^n [y^1
(t),...,y^{n-1}(t);t] \, d t %\nonumber \\  & &
\ + \ \int_{t_0}^t \rho^n_{\ k} [y^1 (t),...,y^{n-1}(t);t] \, d
w^k(t) \ , \label{eq:sol last eq} \end{eqnarray} and is therefore
known. In the language used in the symmetry analysis of
deterministic equations, this can be seen as a ``reconstruction
equation'', and amounts to a (stochastic) quadrature. \EOR

\medskip\noindent
{\bf Example 3.} Let us consider the two-dimensional system (we
write all indices as subscripts to avoid confusion)
\begin{eqnarray}
d y_1 &=& \( e^{y_1} \, - \, \frac12 \, e^{-2 y_1} \) \, d t \ + \ e^{- y_1} \, d w_1 \nonumber \\
d y_2 &=& \frac12 \, e^{y_2} \, \( 2 e^{y_1} \, + \, e^{y_2} \, +
\, e^{2 y_1 +y_2} \) \, d t \ - \ e^{y_1 +y_2} \, d w_1 \ - \
e^{y_2} \, d w_2 \ . \label{eq:example3}  \end{eqnarray} This
admits as symmetry the vector field $ X \ = \ - \, e^{y_2} \, (\pa
/ \pa y_2 )$. Actually, passing to the variables $ x_1 \ = \
e^{y_1}$, $x_2 \ = \ e^{- y_2}$,  the initial system
\eqref{eq:example3} is rewritten as
\begin{eqnarray}
d x_1 &=& x_1^2 \, dt \ + \ d w_1 \nonumber \\
d x_2 &=& - x_1 \, dt \ + \ x_1 \, dw_1 \ + \ d w_2 \ . \end{eqnarray}
In these coordinates, $X = (\pa / \pa x_2 )$. \EOR

\section{Multiple symmetries and reductions for systems of SDEs}
\label{sec:multiple}

It is natural to ask now what happens if the system of SDEs
\eqref{eq:Ito} admits more than one symmetry, in particular -- in
view of well known results holding for deterministic equations
\cite{\symmref} -- what happens if it admits a $r$-parameter
solvable group of symmetry. This was stated in \cite{Koz2,Koz3}
but without a formal proof. We will now give a formal statement
and a detailed proof.

\medskip\noindent
{\bf Theorem 3.} {\it Suppose the system \eqref{eq:Ito} admits an
$r$-parameter solvable algebra $\g$ of simple deterministic
symmetries, with generators \beq \mathbf{X}_k=\sum_{i=1}^n
\varphi_k^i(x,t)\frac{\pa}{\pa x_i} \qquad k=1,...,r, \eeq acting
regularly with $r$-dimensional orbits.

Then it can be reduced to a system of $m = (n-r)$
equations, \beq d y^i \ = \ g^i (y^1,...,y^m;t) \, d t \ + \
\s^i_{\ k} (y^1,...,y^m;t) \, d w^k \ \ \ \ (i,k=1,...,m) \eeq and
$r$ ``reconstruction equations'', the solutions of which can be
obtained by quadratures from the solution of the reduced
$(n-r)$-order system.

In particular, if $r=n$, the general
solution of the system can be found by quadratures.}

\medskip\noindent
{\bf Proof.} We want to proceed by induction on the dimension $r$
of the algebra (for $r=1$ this theorem obviously reduces to the
previous one).

First of all we recall that the hypothesis on the algebraic
structure of $\g$ implies that there exists a chain of
algebras $\g^{(k)}$ ($k=1,...,r$) of dimension $k$ such that
$\g^{(0)}\subset\g^{(1)}\subset \dots\subset \g^{(r+1)} = \g$, and
each algebra $\g^{(k)}$ is a normal subalgebra of $\g^{(k+1)}$,
i.e.
$$ [\g^{(k)},\g^{(k+1)}] \subset \g^{(k)} \ .  $$

Suppose now we have an $r=s+1$-parameter solvable group of
symmetries $G$ with Lie algebra $\g$. Let $\mathbf{X}_{s+1}$ be a
generator of $\g^{(s+1)}$ that does not lie in $\g^{(s)}$, and
consider the one-parameter group $G_{s+1}$ generated by it. Thanks
to the solvability condition, the system is invariant under the
action of $G_{s+1}$, since its action coincides to that of the
quotient $\g^{(s+1)}/\g^{(s)}$. Thus we can invoke Theorem 2 and
reduce the system via a change of variable $x \to \wt{x}$, getting
a ``reduced'' system of $n-1$ equations for  $\wt{x}^1
(t),...,\wt{x}^{n-1} (t)$; and a last ``reconstruction'' equation
which amounts to \beq \wt{x}^n=\wt{x}^n(t_0) \ + \ \int_{t_0}^t
\overline{f}^n(\wt{x}^1,...,\wt{x}^{n-1},t) \, dt  \ + \
\int_{t_0}^t \overline{\sigma}^n_k(\wt{x}^1,...,\wt{x}^{n-1},t) \,
dw^k(t) \ . \label{eq:ricostruzione} \eeq

The key observation now is that it has been proven (Lemma 1) that
the symmetries are maintained under changes of variables; thus we
know for sure that the remaining $s=r-1$ symmetries of the
original system are still symmetries of the reduced one (once
expressed in the new variables).

This allows to iterate the previous step and carry on the
procedure described above, using the remaining $s$ vector fields
one by one. After performing all the allowed $r$ steps we will
have a reduced system of $n-r$ equations for $(\wh{x}^1 (t)
,\dots,\wh{x}^{n-r} (t) )$, and $r$ ``reconstruction'' equations
generalizing \eqref{eq:ricostruzione}. Note that if we are able to
obtain a solution to the reduced system, solutions to the full
system require to solve the reconstruction equation in the
``proper'' order, i.e. the one dictated by the solvable structure
of the symmetry algebra (to obtain solutions to the original
equations we will of course also have to invert all the change of
coordinates performed at each step in the reduction procedure).

Note also that if $r=n$, at the last step (after using $r-1$
symmetries) the reduced system will amount to a single equation,
with a symmetry of the form considered in Theorem 1; using this we
will get, as in Theorem 1, a single SDE with coefficients
depending only on the variable $t$, hence an integrable SDE.
Solutions to the full systems are then obtained by solving the
reconstructions equations. \EOP

\medskip\noindent
{\bf Remark 8.} We stress that the solvability hypothesis was
essential to guarantee that the actions of the quotients
$G^{(k+1)}/G^{(k)}$ (where $G^{(k)}$ is the Lie group generated by $\g^{(k)}$, and so on) were symmetries of the system, according to
the fact that if a normal subgroup $H$ of the group $G$ is a
symmetry of a system, then $G$ itself is also a symmetry if and
only if the system reduced under $H$ admits the quotient $G/H$ as
a symmetry group.
\EOR

\medskip\noindent
{\bf Example 4.} We will just refer the reader to Example 4.2 in Kozlov paper \cite{Koz1}. It is shown there that any equation of the form
\begin{eqnarray}
d x_1 &=& (a_1 + b_{11} x_1 + b_{12} x_2 ) \, d t \ + \ s_{11} \, d w_1 \ + \ s_{12} \, d w_2 \nonumber \\
d x_2 &=& (a_2 + b_{21} x_1 + b_{22} x_2 ) \, d t \ + \ s_{21} \,
d w_1 \ + \ s_{22} \, d w_2 \label{eq:example4} \end{eqnarray}
with $a_i$, $b_{ij}$ and $s_{ij}$ real constant and satisfying the
``full rank condition'', which in this case just reads
$$ \mathtt{Det} \begin{pmatrix} s_{11} & s_{12} \\ s_{21} & s_{22} \end{pmatrix} \ \not= \ 0 \ ,  $$
admits symmetries
$$ X \ = \ \vphi_1 (t) \ \frac{\pa}{\pa x_1} \ + \ \vphi_2 (t) \ \frac{\pa}{\pa x_2} \ , $$
with $\vphi_i$ arbitrary smooth functions.
Looking at two vector fields of the form
$$ X_i \ = \ A_i (t) \ \frac{\pa}{\pa x_1} \ + \ B_i (t) \ \frac{\pa}{\pa x_2} \ , $$
with $\Delta := A_1 B_2 - A_2 B_1 \not= 0$ to ensure independence,
the corresponding change of variables is
$$ y_1  \ = \ \frac{B_2 (t) \, x_1 \ - \ A_2 (t) \, x_2}{A_1 (t) \, B_2 (t) \ - \ A_2 (t) \, B_1 (t) } \ ; \ \
y_2  \ = \ \frac{- B_1 (t) \, x_1 \ + \ A_1 (t) \, x_2}{A_1 (t) \, B_2 (t) \ - \ A_2 (t) \, B_1 (t) } \ . $$
With this, the system \eqref{eq:example4} is mapped into
{\small
\begin{eqnarray*}
d y_1 &=& \( \frac{a_1 B_2 - a_2 A_2}{A_1 B_2 - A_2 B_1} \) \, dt \ + \
\( \frac{B_2 s_{11} - A_2 s_{21}}{A_1 B_2 - A_2 B_1} \) \, d w_1 \ + \
\( \frac{B_2 s_{12} - A_2 s_{22}}{A_1 B_2 - A_2 B_1} \) \, d w_2 \\
d y_2 &=& \( \frac{a_2 A_1 - a_1 B_1}{A_1 B_2 - A_2 B_1} \) \, d t \ + \
\( \frac{A_1 s_{21} - B_1 s_{11}}{A_1 B_2 - A_2 B_1} \) \, d w_1 \ + \
\( \frac{A_1 s_{22} - B_1 s_{12}}{A_1 B_2 - A_2 B_1} \) \, d w_2 \ , \end{eqnarray*} }
which is indeed of the form
$$ d y^i  \ = \ F^i (t) \, d t \ + \ S^i_{\ k} (t) \, d w^k $$ and hence integrable.
The reader is referred to \cite{Koz1} for full detail. \EOR

\section{Random maps and Kozlov theorem}
\label{sec:rankoz}

We have so far considered changes of variables of the form
\eqref{eq:smap}; we will now consider a more general class of
transformations, i.e. \emph{random maps} \cite{ArnImk}.

Lemma 1 does still guarantee that symmetries are preserved under
(simple) random maps, but in this case the discussion becomes more
involved, and we will limit to consider simple random maps, see
\eqref{eq:srmap} and \eqref{eq:Xran}, and the framework of
\emph{scalar} Ito equations, i.e. the analogue of the situation
discussed in Section \ref{sec:Kozscalar} (systems will be
discussed in a forthcoming contribution).

The difficulty here lies in that we are not guaranteed a simple
random map \eqref{eq:srmap} will take an Ito equation into an
equation of Ito type.
In fact, let us consider a scalar Ito equation \eqref{eq:koz2} and
assume the existence of a symmetry in the form $X = \vphi (y,t,w)
\pa_y$. If we change coordinates passing to \beql{eq:rr1} x \ = \
\Phi (y,t,w) \ = \ \int \frac{1}{\vphi (y,t,w) } \ d y  \eeq
(this integral is defined up to an additive function $\b
(t,w)$, which is a ``constant of integration'' in this framework),
the vector field $X$ is mapped again into $X = \pa_x$, and Lemma 1
guarantees this is a symmetry of the transformed equation, which
we write as $d x = f d t + \s dw$. Knowing that $X = \pa_x$ is a
symmetry for this equation guarantees -- in view of the
determining equations \eqref{eq:deteqItoR1}, \eqref{eq:deteqItoR2}
-- that $ (\pa f / \pa x) = 0 $, $(\pa \s / \pa x) = 0$; but now
the coefficients $f$ and $\s$ could depend on $w$ as well, in
which case the transformed equation is not even an Ito one.

\medskip\noindent
{\bf Remark 9.} On the other hand, \emph{if} the equation
\eqref{eq:koz2} can be obtained from an integrable equation,
i.e. one with $\wt{f} = \wt{f} (t)$, $\wt{\s} = \wt{\s} (t)$,
via a simple change of coordinates $y = \Theta (x,t,w)$, then
necessarily the map identified by $\Phi$ given above -- or more
precisely by \emph{some} $\Phi$ as above (the one which defines the
change of variable inverse to the one defined by $\Theta$),
i.e. for some choice of the function $\b (t,w)$ playing the role of integration constant -- takes \eqref{eq:koz2} back to the original one, with no
dependence of the coefficients on $w$. \EOR
\bigskip

In this context, we will find useful to consider how a
simple random change of variables
\beql{eq:PhiR} x \ = \ \Phi (y,t;w) \eeq acts on the the general (formal
and) not necessarily Ito equation \beql{eq:dxRw} d x \ = \ f(t,w)
\, d t \ + \ \s (t,w) \, d w \ . \eeq

\medskip\noindent
{\bf Lemma 2.} {\it Under the simple random change of variables
\eqref{eq:PhiR}, the general formal equation \eqref{eq:dxRw}
is mapped into a (formal) equation of the type
\beql{eq:ranw} d y \ = \ \^F (y,t,w) \, d t \
+ \ \^S (y,t,w) \, d w \eeq with coefficients given by
\beql{eq:FSRww} \^F (y,t,w)  \ = \ \frac{f - \Phi_t - (1/2) \Delta
\Phi }{\Phi_y} \ , \ \ \^S (y,t,w) \ = \ \frac{\s -
\Phi_w}{\Phi_y} \ . \eeq}

\medskip\noindent
{\bf Proof.} This follows by a standard computation. We start from
\eqref{eq:dxRw}, and operate with the general change of variables
\eqref{eq:PhiR}; note that $t$ and $w$ -- and therefore also $f (t,w)$
and $\s (t,w)$ -- are not changed in any way, and that to have a proper
change of variables we must require (we recall that here $\Theta$ denotes the change of coordinates inverse to $\Phi$) $\Phi_y \not= 0$ $\forall y$ and $\Theta_x \not= 0$ $\forall x$.

Using Ito formula, \begin{eqnarray} d x &=& \Phi_y \, d y \ + \
\Phi_t \, dt \ + \ \Phi_w \, d w \ + \ (1/2) \, \Delta (\Phi) \,
d t \nonumber \\
&=& \Phi_y \, d y \ + \ \Phi_t \, dt \ + \ \Phi_w \, d w \ + \
(1/2) \, \( \Phi_{yy} \, \^S^2 \, + \, \Phi_{ww} \, + \, 2 \,
\^S \, \Phi_{yw} \) \, d t \ ; \label{eq:ran11} \end{eqnarray}
note that here we have introduced $\^S$, i.e. the coefficient of
the noise term in the equation for $y$, which we have not yet
determined. From \eqref{eq:ran11} and \eqref{eq:dxRw} we have \beq
\Phi_y \, dy \ = \ \[ f  - \Phi_t - (1/2) \Delta (\Phi ) \] \, d t
\ + \ \[ \s - \Phi_w \] \, d w \ ; \eeq recalling that $\Phi_y$ is
never vanishing, we divide by $\Phi_y$ and obtain an equation of
the form \eqref{eq:ranw} with coefficients given precisely by
\eqref{eq:FSRww}. \EOP

\medskip\noindent
{\bf Corollary 1.} {\it The equation \eqref{eq:ranw} is obtained
from an equation \eqref{eq:dxRw} by a simple change of coordinates
\eqref{eq:Phi} if and only if there exist functions $f$, $\s$ and
$\Phi$ satisfying the equations
\beql{eq:corollary1} \( \s \ - \
\Phi_w \) \ = \ \^S \ \Phi_y \ , \ \ \ \
\( f \ - \ \Phi_t \ - \ (1/2) \Delta \Phi \) \ = \ \^F \
\Phi_y \ . \eeq}

\medskip\noindent
{\bf Proof.} This is a trivial restatement of Lemma 2 above. \EOP

\bigskip\noindent
{\bf Remark 10.} Obviously eq. \eqref{eq:dxRw} admits $X = \pa_x$
as a Lie-point symmetry. By Lemma 1, this will also be a symmetry
of the transformed equation; on the other hand, in the new
coordinates we have $\pa_x = (\pa y / \pa_x ) \pa_y = (1/\Phi_y) \pa_y$.
Thus, in the situation considered by Lemma 2, the transformed
equation \eqref{eq:ranw} admits the vector field
\beql{eq:rem10} X \ = \ \[ \Phi_y (y,t,w) \]^{-1} \, \pa_y \ := \
\varphi (y,t,w) \, \pa_y \eeq as a (simple) Lie-point symmetry, as
follows at once from Lemma 1.

This can also be checked
by direct computation using the explicit form of \eqref{eq:deteqItoR1}, \eqref{eq:deteqItoR2}. \EOR
\bigskip

We are now interested in the case where the equation
\eqref{eq:ranw} is actually an Ito equation,
\beql{eq:ranIto} \^F (y,t,w) \ = \ F(y,t) \ , \ \ \ \^S (y,t,w) \
= \ S(y,t) \ . \eeq

\medskip\noindent
{\bf Lemma 3.} {\it Let the equation \eqref{eq:ranw} be of Ito
type, i.e. let \eqref{eq:ranIto} be satisfied. Then the functions
$f,\s,\Phi$ of Lemma 2 satisfy the equations \beql{eq:LR3} \s \ =
\ \Phi_w \ + \ S \, \Phi_y \ ; \ \ \ \ \ f \ = \ \Phi_t \ + \ F \,
\Phi_y \ + \ (1/2) \, \( \Phi_{ww} \, + \, 2 \, S \, \Phi_{yw} \, + \, S^2 \, \Phi_{yy} \) \ . \eeq}

\medskip\noindent
{\bf Proof.} This is exactly \eqref{eq:corollary1} with $S$ taking
the place of $\^S$ and writing explicitly $\Delta \Phi$. \EOP

\medskip\noindent
{\bf Lemma 4.} {\it Let the ``target'' equation \eqref{eq:ranw} be of Ito
type, i.e. let \eqref{eq:ranIto} hold; moreover,
let \eqref{eq:ranw} be obtained from \eqref{eq:dxRw} by a simple random map
\eqref{eq:PhiR}. Then the ``source'' equation \eqref{eq:dxRw} is of Ito type
if and only if
\beql{eq:dsfw0} \Phi_{ww} \ + \ S \, \Phi_{yw} \ = \ 0 \ ; \ \ \ \ \
\Phi_{tw} \ + \ F \, \Phi_{yw} \ + \ (1/2) \, \( \Delta \Phi \)_w \ = \ 0 \ . \eeq}

\medskip\noindent
{\bf Proof.} We are in the framework of Lemma 3. The equation
\eqref{eq:dxRw} is of Ito type if and only if the coefficients $f$
and $\s$ do not depend on $w$. The requirement that $\s_w = 0$ and
$f_w = 0$, in view of \eqref{eq:LR3}, are exactly the first and second equation respectively in \eqref{eq:dsfw0}. \EOP

\medskip\noindent
{\bf Remark 11.} Introducing $\Gamma := \Phi_w$, the equations
\eqref{eq:dsfw0} are rewritten as
\beq \Gamma_{w} \ + \ S \, \Gamma_{y} \ = \ 0 \ ; \ \ \ \
\Gamma_{t} \ + \ F \, \Gamma_{y} \ + \ (1/2) \, \( \Delta \Gamma \)
\ = \ 0 \ . \label{eq:dfsw0chi} \eeq (To obtain these, recall that now by
assumption $S_w = 0$). \EOR
\bigskip

\medskip\noindent
{\bf Theorem 4.} {\it Let the Ito equation \beql{eq:dyRR} d y \ =
\ F(y,t) \, d t \ + \ S (y,t) \, d w \eeq admit as Lie-point
symmetry the simple random vector field \beql{eq:Xrvf} X \ = \vphi (y,t,w) \
\pa_y \ . \eeq
If there is a determination of \beql{eq:vphiPhi}
\Phi (y,t,w) \ = \ \int \frac{1}{\vphi (y,t,w)} \ d y \eeq such
that the equations \eqref{eq:dsfw0} are satisfied,
then the equation is reduced to the explicitly integrable form
\beql{eq:dxRint} d x \ = \ f(t) \, d t \ + \ \s (t) \, d w \eeq by
passing to the variable $x = \Phi (y,t,w)$.}

\medskip\noindent
{\bf Proof.} When we operate with the map  $x = \Phi (y,t,w)$, the equation for $x$ obtained from \eqref{eq:dyRR} will be written as $d x = f d t + \s dw$.

With such a map -- for \emph{any} determination of $\Phi$ -- the
vector field will $X$ will just read as $X = \pa_x$ in the new coordinates,
and we are guaranteed by Lemma 1 this will be a symmetry of the transformed equation. By \eqref{eq:deteqItoR1}, \eqref{eq:deteqItoR2}, this means that $f_x = 0$, $\s_x = 0$, i.e. the equation in the $x$ coordinate will be of the form
\eqref{eq:dxRw}.

Moreover, by Lemma 4, the coefficients of this satisfy $f_w = 0 = \s_w$,
i.e. we have a proper Ito equation, if and only if $\Phi$
satisfies the equations \eqref{eq:dsfw0}, in which
now $F$ and $S$ are assigned functions, i.e. the coefficients of
our equation \eqref{eq:dyRR}.
Thus, by choosing such a $\Phi$ (among
those of the form \eqref{eq:vphiPhi}), we are guaranteed the $f$
and $\s$ do not depend on $x$ nor on $w$, i.e. the equation is
precisely of the form \eqref{eq:dxRint}. \EOP
\bigskip

\medskip\noindent
{\bf Remark 12.} This Theorem is based on the existence of a
determination of an integral with certain properties. It is
natural to wonder how one goes in order to study the existence of
such a determination. In other words, one would like to have a
criterion based on the directly available data, i.e. the functions
$F(y,t)$, $S(y,t)$ and $\vphi (y,t,w)$. This is provided by the next Theorem. \EOR

\medskip\noindent
{\bf Theorem 5.} {\it Let the Ito equation \eqref{eq:dyRR} admit
as Lie-point symmetry the simple random vector field \eqref{eq:Xrvf};
define $\ga (y,t,w) := \pa_w ( 1 / \vphi )$.

If the functions $F(y,t)$, $S(y,t)$ and $\ga (y,t,w)$ satisfy the
relation \beql{eq:bcomp} S \, \ga_{t} \ + \ S_t \, \ga \ = \ F \,
\ga_{w} \ + \ (1/2) \, \[ S \, \ga_{ww} \ + \ S^2 \, \ga_{yw} \]
\ , \eeq then the equation \eqref{eq:dyRR} can be mapped into an
integrable Ito equation \eqref{eq:dxRint} by a simple random
change of variables.}

\medskip\noindent
{\bf Proof.} For any given determination $\Xi$ of the integral, i.e. any function satisfying $ \Xi_y = (1/\vphi )$, the most general integral is given by
$$ \Phi \ = \ \int \frac{1}{\vphi} \, d y \ = \ \Xi (y,t,w) \ -
\ \a (t,w) \ , $$ with $\a$ an arbitrary function. With this, we
write
$$ \Gamma \ = \ \Phi_w \ = \ \Xi_w \ - \ \a_w \ := \ \psi \ - \ \b \
. $$ The equations \eqref{eq:dfsw0chi} read then
\beq \b_w \ = \ \psi_{w} \ + \ S \, \psi_{y} \ , \ \ \ \
\b_t \ = \ \psi_{t} \ + \ F \, \psi_{y} \ + \ (1/2) \, \[ \(
\Delta \psi \) \ - \ \Delta (\b ) \] \ . \label{eq:dsfw0b0} \eeq
Noting now that $\Delta \b = \b_{ww}$, and that in view of the first of
\eqref{eq:dsfw0b0} we can write $\b_{ww} = \psi_{ww} + S
\psi_{yw}$, we rewrite the equations as
\beq \b_w \ = \ \psi_{w} \ + \ S \, \psi_{y} \ , \ \ \ \
\b_t \ = \ \psi_{t} \ + \ F \, \psi_{y} \ + \ (1/2) \, \[ S \,
\psi_{yw} \ + \ S^2 \, \psi_{yy} \] \ . \label{eq:dsfw0b1}
\eeq
The condition of compatibility $\b_{tw} = \b_{wt}$ reads therefore
\beql{eq:bcomp0} S \, \psi_{yt} \ + \ S_t \, \psi_{y} \ = \ F \,
\psi_{yw} \ + \ (1/2) \, \[ S \, \psi_{yww} \ + \ S^2 \,
\psi_{yyw} \] \ . \eeq

This can be slightly simplified by writing $ \ga
 := \psi_y$; actually, recalling that $\psi =
\Xi_w$, and that $\b_y = 0$, this definition yields \beql{eq:ga2} \ga (y,t,w) \ = \ \psi_y \ = \ \Xi_{wy} \ = \ \Phi_{wy} \ = \ \pa_w \(
1 / \vphi \) \ , \eeq i.e. exactly the definition provided in the statement.

Thus the compatibility condition \eqref{eq:bcomp0} can
be written entirely in terms of the accessible data, i.e. the
coefficients $F$ and $S$ in \eqref{eq:dyRR} and the coefficient
$\vphi$ of the symmetry vector field. More precisely, plugging
$\ga = \psi_y$ into \eqref{eq:bcomp0}, we get exactly
\eqref{eq:bcomp}. \EOP

\medskip\noindent
{\bf Remark 13.} Note that for $\vphi$ independent of $w$ (i.e.
deterministic simple symmetries), which entails $\ga = 0$, the
equation \eqref{eq:bcomp} is always trivially satisfied -- as it
should be. \EOR

\medskip\noindent
{\bf Remark 14.} Even in the case of deterministic symmetries, it
can make sense to consider simple random changes of variables: by
enlarging the set of considered transformations, we could obtain a
simpler transformed equation (this will be the case in Example 5).
\EOR
\bigskip

The Theorems 4 and 5 identify the presence of a simple random symmetry \eqref{eq:Xrvf} such that the compatibility condition \eqref{eq:bcomp} is satisfied as a \emph{sufficient} condition for integrability. It is quite simple to observe this is also a \emph{necessary} condition.

\medskip\noindent
{\bf Theorem 6.} {\it Let the Ito equation \eqref{eq:dyRR} be reducible to the integrable form \eqref{eq:dxRint} by a simple random change of variables \eqref{eq:PhiR}. Then necessarily \eqref{eq:dyRR} admits \eqref{eq:rem10} as a symmetry vector field, and -- with $\ga = \pa_w (1/\vphi)$ -- \eqref{eq:bcomp} is satisfied.}

\medskip\noindent
{\bf Proof.} Let $y=\Theta(x,t,w)$ be the change of variables inverse to \eqref{eq:PhiR}. By assumption, there is an integrable Ito equation \eqref{eq:dxRint} which is mapped into \eqref{eq:dyRR} by $\Theta$. Equation \eqref{eq:dxRint} admits $X = \pa_x$ as a symmetry, and by Lemma 1 this will also be a symmetry of \eqref{eq:dyRR}; on the other hand, $X$ is written in the $(y,t,w)$ coordinates exactly in the form \eqref{eq:rem10}. The fact that \eqref{eq:bcomp} is satisfied follows at once from the assumption that both \eqref{eq:dxRint} and \eqref{eq:dyRR} are Ito equations. \EOP
\bigskip

Thus, in the end, even within the framework of (simple) random maps and (simple) random symmetries, the presence of a simple random symmetries -- now with the additional requirement that a compatibility condition \eqref{eq:bcomp} is satisfied -- is a \emph{necessary and sufficient} condition for a given (scalar) Ito equation to be integrable by a (simple random) change of variables.

\section{Examples of integration via random symmetries}
\label{sec:exarandom}

In this Section we will present several examples of reduction to integrable form via (changes of variables identified by) simple random symmetries. As this kind of reduction is new in the literature, we will discuss these examples in greater detail than the previous ones.

\medskip\noindent
{\bf Example 5.} Let us consider the equation \beql{eq:example5} d
y \ = \ \( (1/2) \, t^2 \, e^{-t} \ - \ y \) \, d t \ + \ (t \,
+ \, k ) \, e^{- t} \, d w \ ; \eeq this admits as symmetry $X =
e^{- t} \pa_y $. Here $\vphi_w = 0$ and the equation
\eqref{eq:bcomp} is trivially satisfied. According to our general
result, we should therefore use the change of variable described
by \beql{eq:phigenexample5} \Phi \ = \ \int e^t \, d y \ = \ e^t
\, y \ + \ \b (t,w) \ ; \eeq with this, the equation
\eqref{eq:example5} is mapped into \beq d x \ = \ \( \frac{t^2}{2}
\, + \, (1/2) \, \b_{ww} \, + \, \b_t \) \, d t \ + \ \( k \, +
\, t \, + \, \b_w \) \, d w \ . \eeq As expected, its coefficients
are always independent of $y$.

For this to be an Ito equation, from the coefficient of $d w$ it
is needed to have $\b_{ww} = 0$, hence $\b (t,w) = b (t) + b^{(1)}
(t) w $, and from the coefficient of $d t$ we get that actually we
should require $b^{(1)} (t) = c$. In other words, we have \beq
\Phi \ = \ c \, w \ + \ e^t \, y \ + \ b (t) \ , \eeq where the
constant  $c$ and the function $b(t)$ are arbitrary. With the map
$x = \Phi(y,t,w)$ and this choice of $\Phi$, we get the equation
\beq d x \ = \ \( \frac{t^2}{2} \ + \ b' (t) \) \, d t \ + \ \( k
\, + \, c \, + \, t \) \, d w \ := \ f(t) \, d t \ + \ \s (t) \, d
w ; \eeq this is always (that is, for any choice of $b(t)$ and $c$)
integrable.

We can get a somewhat simpler equation choosing $c
= - k$ and $b(t) = 0$, which corresponds to $\Phi (y,t,w)  =
 -  k  w +  e^t y$; now the transformed equation
is\footnote{An even more radical simplification is obtained by choosing
$c = -k$ and $b(t) = - t^3/6$, i.e. $\Phi = - k w + e^t y - t^3/6$; with this choice the transformed equation is $d x = t dw$.} \beq d x \ = \ \frac{t^2}{2} \, d t \ + \  t  \, d w \ . \eeq

Note that in this case the symmetry is actually deterministic; if
we were working in the framework of deterministic simple maps, the
only difference would have been that we should have just
considered $c=0$ and $\b (t,w) = b(t)$. \EOR

\medskip\noindent
{\bf Example 6.} The equation
\beql{eq:example6} d y \ = \ - \( e^{- y} \, + \, (1/2) \, e^{-
2 y} \) \, d t \ + \ e^{- y} \, d w \eeq admits the vector field
\beq X \ = \ \frac{e^{-y}}{t+e^y-w+1} \, \pa_y \ := \ \vphi
(y,t,w) \, \pa_y \eeq as symmetry (this is a genuine random
symmetry); condition \eqref{eq:bcomp} is satisfied. We hence have
\beq \Phi \ = \ \int \frac{1}{\vphi (y,t,w) } \ d y \ = \ e^y
(t-w+1)+\frac{e^{2 y}}{2} \ + \ \b (t,w) \ . \eeq

By passing to the variable $x = \Phi (y,t,w)$ the equation
\eqref{eq:example6} reads
$$ d x \ = \ \( \b_t \, + \, (1/2) \, \b_{ww} \, + \, w \, - \, t \, -
\, \frac32 \) \, d t \ + \ \( \b_w \, - \, w \, + \, t \, + \, 1
\) \, d w \ := \ f \, d t \ + \ \s \, d w \ . $$ Again the
coefficients are -- as expected -- always independent of $y$.

In order to have $\s_w = 0$ we must require $\b_{ww} = 1$, i.e.
$\b (t,w) = b (t) + [b^{(1)}(t)]  w  +  (1/2) w^2 $; with this,
requiring $f_w = 0$ enforces $b^{(1)} (t) = c_1 - t$. The equation
reads now \beq d x \ = \ - \, \( 1 \, + \, t \, + \, b' (t) \) \,
d t \ + \ (1 + c_1) \, d w \ , \eeq which is obviously integrable.
Note that with our choice for $\b$ we have got \beq \Phi (y,t,w) \
= \ \frac{w^2}{2} \ + \ \left( c_1 \, - \, e^y \, - \, t \right)
\, w \ + \ \[ e^y \, + \, \frac{e^{2 y}}{2} \, + \, e^y \,  t \, +
\, b (t) \] \ . \eeq

By choosing $c_1= 0$, $b(t) = 0$ we get slightly simpler
expressions: the transformed equation is \beq d x \ = \ - \, \( 1
\, + \, t \) \, d t \ + \  d w \ , \eeq and the integrating map is
\beq \Phi (y,t,w) \ = \ \frac{w^2}{2} \ - \ \left( e^y \, + \, t
\right) \, w \ + \ \[ 1 \, + \, \frac{e^y}{2} \, + \, t \] \, e^y
\ . \eeq

Note that \eqref{eq:example6} also admits a simple \emph{deterministic} symmetry, i.e. $X_0= e^{-y} \pa_y $.\footnote{More generally, any function
$\vphi (y,t,w) = e^{- y} \eta ( e^y + t -w) $ identifies a Lie-point symmetry for \eqref{eq:example6}; see also the following Example 7 and Appendix B in this respect.} \EOR

\medskip\noindent
{\bf Example 7.} Consider the Ito equations
\beql{eq:example7} d y \ = \ y \, p(t) \, d t \ + \ y \, d w \ , \eeq
for $p(u)$ any smooth function. The determining equation \eqref{eq:deteqItoR2} reads in this case
$\vphi = \vphi_w + y  \vphi_y $ and hence yields
$\vphi (y,t,w) = y  \psi (z,t)$, where $z := y  e^{-w}$; with this the other determining equation \eqref{eq:deteqItoR1} reads
$\psi_t + z  [ p(t)  -  (1/2) ]  \psi_z = 0 $ and yields
$ \psi (z,t)  = \eta (\zeta)$, where we have defined \beq \zeta \ = \ z \ \exp \[ R(t) \] \ = \ y \ \exp \[ - w \ + \ R(t) \] \ ; \ \ \
R(t) \ := \ \int_0^t \( \frac12 \, - \, p(u) \) \ d u \ . \eeq Thus, in the end, the simple symmetries of \eqref{eq:example7} are identified by
\beq \vphi (y,t,w) \ = \ y \ \eta (\zeta ) \ , \eeq with $\eta$ an arbitrary  non-zero function. The compatibility condition \eqref{eq:bcomp} is always satisfied.

The change of variables identified by such symmetries are
\beql{eq:Phiex7} x \ = \ \Phi (y,t,w) \ = \ \int \frac{1}{y \ \eta (\zeta)} \ d y \ + \ \b (t,w) \ . \eeq
In order to apply Lemma 3, and in particular \eqref{eq:LR3}, we compute partial derivatives of $\Phi$; in doing this we should distinguish the case $\eta = K$ (with $K \not= 0$ a constant) and the generic one.

In fact, if $\eta' (\zeta ) \not= 0$, we have
\begin{eqnarray*} \Phi_t &=& \b_t \, - \, R' (t) \ e^{- w + R(t)} \ \int (\eta_\zeta / \eta^2 ) \, d y \ = \ \b_t \, - \, R' (t) \ \int (\eta_\zeta / \eta^2) \, d \zeta \ = \ \b_t \, + \, \frac{R' (t)}{\eta} \ ,  \\
\Phi_w &=& \b_w \, + \, e^{- w + R(t)} \ \int (\eta_\zeta / \eta^2 ) \, d y \ = \ \b_w \, + \, \int (\eta_\zeta / \eta^2) \, d \zeta \ = \ \b_w \, - \, \frac{1}{\eta} \ ;  \end{eqnarray*}
by the same computation it is clear that in the case $\eta = K$, and hence $\eta_\zeta = 0$, we get
$$ \Phi_t \ = \ \b_t \ , \ \ \ \Phi_w \ = \ \b_w \ . $$
Obviously we always have $\Phi_y = 1/(y \eta)$.

As for the second order partial derivatives, these are (these formulas also hold for $\eta = K$, simply setting $\eta_\zeta = 0$ in them)
$$ \Phi_{yy} \ = \ - \frac{\eta \ + \ \zeta \, \eta_\zeta}{y^2 \ \eta^2} \ , \ \ \Phi_{ww} \ = \ \b_{ww} \ - \ \frac{\zeta \, \eta_\zeta}{\eta^2} \ , \ \
\Phi_{wy} \ = \ \frac{\zeta \, \eta_\zeta}{y \, \eta^2} \ ; $$
With these -- and recalling $S=y$, $F = y p(t)$, see \eqref{eq:example7} -- we always get $\Delta (\Phi) = \b_{ww}  - (1/\eta )$.
We can now apply \eqref{eq:LR3} to get $f$ and $\s$ of the transformed equation.

\bigskip\noindent
{\bf (A)} {\it ($\eta$ not constant).}
For $\eta \not= K$, applying the first of \eqref{eq:LR3} we get immediately
\beql{eq:sigma7} \s \ = \ \b_w \ . \eeq Applying the second of \eqref{eq:LR3}, for $\eta$ non constant we get $ f = [- 1 + 2 p(t) + 2  R'(t) + \eta \b_{ww} +  2  \eta  \b_t]/(2 \eta)$;  recalling now the expression for $R(t)$, hence $R' (t) = (1/2) - p(t)$, this just reduces to
\beql{eq:effe7} f \ = \ \b_t \ + \ \frac12 \, \b_{ww} \ . \eeq
If now we require that $f$ and $\s$ are both independent of $w$, we get
$$ \b (t,w) \ = \ b(t) \ + \ c \, w \ . $$
With this -- recalling \eqref{eq:sigma7} and \eqref{eq:effe7} -- we obtain\footnote{Note that in this case we obtain $d x = d \beta$; indeed it follows directly from eq.\eqref{eq:LR3} that choosing $\Phi$ to be just the integral part of \eqref{eq:Phiex7} (i.e. setting $b(t,w)=0$)  we get $f(x,t) = \s (x,t) = 0$. Note also this is not the case in Example 6.}
\beql{eq:red7gen} d x \ = \ [b' (t) ] \, d t \ + \ c \, d w \ . \eeq

This shows that we can have several random symmetries -- actually, in this case, infinitely many ones, depending on the choice of an arbitrary non-constant smooth function $\eta$ -- leading to \emph{the same} integrable equation; this situation is discussed in more detail in \ref{sec:appB}.

\bigskip\noindent
{\bf (B)} {\it ($\eta$ constant).}
For $\eta$ constant, applying the first of \eqref{eq:LR3} we get
\beql{eq:sigma7C} \s \ = \ \b_w \ + \ K^{-1} \ ; \eeq applying now also the second of \eqref{eq:LR3} we get
\beql{eq:effe7C} f \ = \ \( b_t \ + \ \frac12 \b_{ww} \) \ + \ \( p(t) \, - \, \frac12 \) \ K^{-1} \ . \eeq
In order to have $s_w = 0$ and $f_w = 0$ we must again require $\b = b(t) + c w $.
Thus for $\eta = K$ the reduced equation is
\beql{eq:red7con} d x \ = \ \frac{1}{K} \ \[ \( p(t) \, - \, \frac12 \ + \ K \, b' (t) \) \ d t \ + \ \( 1 \ + \ K \, c \) d w \] \ . \eeq

Note also that if we choose $\eta$ to be constant, we have a \emph{deterministic}  symmetry; this leads to a \emph{different} reduction (for the same choice of $\b (t,w)$; it is clear that by suitable different choices of $\b$ the two reductions can be made equal), as seen by comparing \eqref{eq:red7gen} and \eqref{eq:red7con}. \EOR

\medskip\noindent
{\bf Example 8.} Consider the Ito equation
\beql{eq:example8} d y \ = \ d t \ + \ y \, d w \ ; \eeq
the only simple Lie-point symmetry this admits is
\beql{eq:Xex8} X \ = \ e^{w - t/2} \ \pa_y \ := \ \vphi (y,t,w) \, \pa_y \ . \eeq
In fact, in this case eq.\eqref{eq:deteqItoR2} yields immediately $$ \vphi (y,t,z) \ = \ y \ \psi (z,t) , \ \ \ \ \ z = y e^{-w} \ ; $$ plugging this into \eqref{eq:deteqItoR1} we obtain $$ \( \psi \ + \ z \, \psi_z \) \ + \ \( \psi_t \, - \, (1/2) \, z \, \psi_z \) \, y \ = \ 0 \ . $$ Each of the two terms in brackets must vanish, hence $\psi (z,t) = \eta(t) / z$ and $\eta = k e^{- t/2}$. Going back to the original variables and function, we obtain \eqref{eq:Xex8}.

For this form of $\vphi$, we have  $\ga = \pa_w (1/\vphi) = e^{t/2 - w}$, and it is easy to check that the compatibility condition \eqref{eq:bcomp} is \emph{not} satisfied.

Let us now consider the changes of variables \eqref{eq:PhiR} such that $X = \pa_x$ in the new variables, i.e. yielding equations of the form \eqref{eq:dxRw}; we know these are of the form
$$ \Phi (y,t,w) \ = \ \int \frac{1}{\vphi} d y \ = \ \( e^{t/2 -w} \) \, y \ + \ \b (t,w) \ . $$
The coefficients $f, \s$ in \eqref{eq:dxRw} are given in this case by \eqref{eq:LR3}, and we obtain
\beq \s \ = \ \b_w \ , \ \ \ \ f \ = \ e^{t/2 - w} \ + \ \frac12 \, \b_{ww} \ + \ \b_t \ ; \eeq requiring $\s_w = 0$ would enforce $\b_{ww} = 0$, i.e.
$$ \b (t,w) \ = \ b_0 (t) \ + \ b_1 (t) \, w \ ; $$ with this, we get $f = e^{t/2 - w} + \b_0' + w b_1'$, and the requirement $f_w = 0$ amounts to
$$ b_1' (t) \ = \ e^{t/2 - w} \ , $$ which does not admit any solution.

Thus in this case the equation admits a simple random symmetry, but it can \emph{not} be transformed into an integrable Ito equation by a simple random map; this corresponds to the fact that the compatibility condition \eqref{eq:bcomp} is \emph{not} satisfied. \EOR

\section{Conclusions and discussion}

We have considered in constructive terms the relation between
\emph{symmetry} and \emph{integrability} -- or at least
\emph{reducibility} -- for stochastic differential equations (or
systems thereof).

We started by recalling that symmetries of an Ito equation are
defined only algebraically, not geometrically;  thus -- contrary
to the deterministic case or even to the Stratonovich one -- in
this context it is not granted that symmetry are preserved under a
change of variables. On the other hand, preservation of
\emph{simple deterministic} symmetries (which are exactly the
symmetries relevant in Kozlov theory \cite{Koz1,Koz2,Koz3}) of Ito
equations is guaranteed, as can be shown by considering the
relation with symmetries of the associated Stratonovich equations
\cite{GLjnmp,Unal}.

We have then considered Kozlov results \cite{Koz1,Koz2,Koz3}; our
Theorems 1, 2 and 3 reproduce them,  but while Kozlov was
interested in providing \emph{sufficient} symmetry conditions for
integrability or reducibility, we have also considered
\emph{necessary} ones; actually showing that Kozlov conditions are
\emph{necessary and sufficient}. Moreover, we have provided fully
explicit and complete proofs to his results.

On the other hand, one can consider more general sets of maps and
hence symmetries; in particular, we have considered \emph{simple
random maps and symmetries} \cite{GS}. In this case the
preservation of symmetries of an Ito equation is granted by a
recent result of ours \cite{GLjnmp}.

We have then tackled, confining our study to the scalar case, the
problem of extending Kozlov theory to the framework of (simple) random maps and symmetries. This extension is provided by Theorem 5 and
Theorem 6, which are a direct extension of Kozlov theorem
\cite{Koz1} and of our completion of it (Theorem 1), except that now we have to ask further conditions \eqref{eq:dsfw0} to ensure we obtain an Ito
equation, or equivalently a compatibility condition \eqref{eq:bcomp} to be sure they can be satisfied.

Some final considerations are in order here, also regarding
possible further extensions of Kozlov theory.

\medskip\par\noindent
{\tt (a)} Our study of random maps and possibly random symmetries
was limited to scalar equations; we trust it can be extended to
the case of systems, but such study is deferred to a forthcoming
contribution.

\medskip\par\noindent
{\tt (b)} A ``naive'' extension of Kozlov theory would consider
reduction to SDEs which can be integrated albeit not being of the
form considered here (and hence not ``immediately''integrable);
this would be e.g. the case for separable SDEs. As mentioned
above, in this case one does not obtain anything new w.r.t. the
standard Kozlov theory; this is discussed in detail in \ref{sec:extended}.

\medskip\par\noindent
{\tt (c)} Further generalizations can be envisaged, albeit they
have not been considered here. The discussion by Kozlov
\cite{Koz1,Koz2,Koz3} suggests that there is no point in
considering deterministic symmetries which are not simple, i.e.
which also act on $t$; on the other hand, there is probably a
point in considering W-symmetries \cite{GRQ2}, which will be done
elsewhere.

\medskip\par\noindent
{\tt (d)} Also, here we have mainly investigated the relation
between symmetries and integrability, which corresponds to an
extremely simplified form of SDE; but other -- less radical --
simplification of the SDEs under study can also be of interest.
One has been considered in Section \ref{sec:Kozsystems}, where we
considered systems which are reducible to a system of lower
dimension, albeit not integrable, plus some ``reconstruction
equations''. One could also consider e.g. systems which are
separable into two or more subsystems, with no
``reconstruction equation''. Such systems can also be
characterized, albeit less immediately, in symmetry terms and it
is thus possible that systems which can be brought into such form
can also be detected in terms of their symmetry properties.

\medskip\par\noindent
{\tt (e)} In this note we have only considered Ito equations. One
could of course consider also more general stochastic differential
equations and their symmetries (see e.g. the recent paper
\cite{ADVMU}; some of its authors also propose an alternative
approach to reduction and reconstruction of stochastic
differential equations via symmetries \cite{DVMU}). In this
framework the approach presented here can be of use to identify
such more general equations which can be mapped back to an Ito
integrable equation.

\medskip\par\noindent
{\tt (f)} Finally we note that, as in the deterministic case, one
expects that the relation between symmetries and integrability
displays special features in the case of stochastic equations with
a \emph{variational} origin. This lies beyond the limit of the
present study but several results exist in this direction, see
e.g. \cite{VSR5,VSR3,VSR4,VSR1,VSR2}.

%\vfill

\section*{Acknowledgements}

We are most grateful to an unknown Referee of the companion paper \cite{GLjnmp} who pointed out a missing term in the formula of the Ito Laplacian \eqref{eq:Delta}, thus avoiding incorrect conclusions. We are equally grateful to the Referee of the present paper for detecting several errors and for several other constructive remarks; and above all for kindly but firmly pushing us to substantially improve the paper. We are also grateful to the Editor for his patience in awaiting the revised version.
The first part of this work follows from the M.Sc. Thesis of CL. We thank Francesco Spadaro (EPFL Lausanne) for useful discussions. A substantial part of this work was performed while GG was visiting SMRI; the work of GG is
also supported by GNFM-INdAM.

%\vfill

%\newpage

\begin{appendix}

\section{The ``extended'' Kozlov theorem}
\label{sec:extended}

Equations of the form \eqref{eq:koz1} are not the only SDEs which can be integrated. Another class is provided by equations of the form
$$ d x \ = \ \b (x) \, f(t) \ d t \ + \ \b (x) \, \s (t) \ d w \ . \eqno(A.1) $$
Note that these are integrated to provide the solution in implicit form only (in general), i.e. as
$$ B(x) \ := \ \int \frac{1}{\b (x)} \, d x \ = \ \int f(t) \, d t \ + \ \int \s (t) \, d w \ . $$

Correspondingly we would have an extension of Theorem 1 to
consider this case. It turns out, however, that equations of the
form (A.1) do not in general admit simple Lie symmetries. This is
detailed in Lemma A.1 below.

\medskip\noindent
{\bf Lemma A.1.} {\it The equation (A.1) does not admit
any simple Lie point symmetry generator unless $\b (x) = x$; in
this case the admitted symmetry generator is $$ X \ =
\ x \, \pa_x \ . \eqno(A.2) $$}

\medskip\noindent
{\bf Proof.} The equation \eqref{eq:deteqIto2} reads in this case
$$ \s \, \( \b \, \vphi_x \ - \ \vphi \, \b_x \) \ = \ 0 \ , $$ and
hence (for $\s \not= 0$) has solution $$ \vphi (x,t) \ = \ A(t) \
\b (x) \ . $$ Plugging this general expression for $\vphi$ into
\eqref{eq:deteqIto1}, we get $$ A' (t) \ \b (x) \ = \ - \, (1/2)
\ [ \b (x) \, \s (t) ]^2 \ A(t) \, \b_{xx} (x) \ . $$ This in turn
implies $$ \frac{A'}{A \, \s} \ = \ - (1/2) \b \b_{xx} \ ; $$
but here the expression on the l.h.s. is a function of $t$ alone,
that on the r.h.s. of $x$ alone. Hence we must have $$ \frac{A'}{A
\, \s} \ = \ c_0 \ = \  - (1/2) \b \b_{xx} \ , $$ with $c$ a
numerical constant.

The first of these equations yields $$ A(t) \ = \ c_1 \ \exp
\left[ c_0 \ \int g^2 (t) \ d t \right] $$ with $c_1$ an arbitrary
constant; the second one requires $$  \b \ \b_{xx} \ = \ - 2 c_0 \
, $$ and the solution is given by $$ \b (x) \ = \ b_0 \ x \ ,
\eqno(A.3) $$  with $b_0$ another constant. This also implies
$c_0= 0$, and hence $A(t) = c_1$.

In other words, (A.1) has a \emph{simple} Lie
symmetry\footnote{Equations with different $\b (x)$ could in
principles admit more general symmetry generators, i.e. could have
symmetries which are not simple ones. We will not discuss these
here.} if and only if $\b (x)$ is of the form (A.3)
(then we can move the constant $b_0$ into the $f(t)$ and $g(t)$
functions, i.e. we can assume $b_0 = 1$ with no loss of
generality). \EOP
\bigskip

It should be noted that for $\b (x)$ of the form (A.3) and the
associated (A.1), the determining equations give (A.2). The
symmetry group generated by (A.2) is just a scaling one, $x \to
\lambda x$.

As mentioned above, one could in principles follow the approach of
Theorem 1 and get a result similar to the one there for equations
which can be reduced to the form (A.1). However, we have
seen that a characterization in terms of simple symmetries is
possible only for $\b$ as in (A.3). But for such a case
we can come back to the case considered by Kozlov.

\medskip\noindent
{\bf Lemma A.2.} {\it The equation (A.1) in the case
(A.3) is mapped to an equation of the form
\eqref{eq:koz1} by the change of variables $$ x \ = \ \Phi (y) \
:= \ e^y \ . $$}

\medskip\noindent
{\bf Proof.} By direct computation. \EOP

%\vfill

\section{Multiple simple random symmetries}
\label{sec:appB}

We have observed, in Examples 7 and 8, that one can have a situation where random symmetries correspond to a certain pattern in which enters an arbitrary function. Here we want to discuss this situation -- and the conditions in which it can occur -- in some detail. We will again confine ourselves to scalar Ito equations, as in Sections \ref{sec:rankoz} and \ref{sec:exarandom}.

First of all, given an Ito equation\footnote{We stress this is written in terms of the $x$ variable, and correspondingly the coefficient of the Wiener process in the equation is $\s (x,t)$: in this framework -- and hence in this Appendix -- the Ito Laplacian is written exactly as in \eqref{eq:Delta}.} \eqref{eq:Ito} we introduce the linear operators
\beq L \ := \ \pa_t \ + \ \[ f(x,t)  \ - \ \frac12 \, \s (x,t) \, \s_x (x,t)  \] \, \pa_x  \ ; \ \ \ \ M \ := \ \pa_w \ + \ \s(x,t) \, \pa_x \ . \eeq
With these, the determining equations \eqref{eq:deteqItoR1}, \eqref{eq:deteqItoR2} are written as
\begin{eqnarray}
L (\vphi ) \ + \ \frac12 \, M^2 (\vphi ) &=& \vphi \, (\pa_x F) \ , \nonumber \\
M (\vphi) &=& \vphi \, (\pa_x \s ) \ . \label{eq:Bde1} \end{eqnarray}

Suppose now that we have a solution $\chi(x,t,w)$ to these equations, and let us look for solutions in the form $\vphi = \psi \chi$.
Plugging this into \eqref{eq:Bde1}, we obtain
\begin{eqnarray}
\chi \, L(\psi) \ + \ \psi \, L(\chi) \ + \ \frac12 \ \[ \chi \, M^2 (\psi) \ + \ 2 \, M(\psi ) \, M(\chi ) \ + \ \psi \, M^2 (\chi) \] &=& \psi \, \chi \, (\pa_x F) \ , \nonumber \\
\chi \, M(\psi) \ + \ \psi \, M(\chi) &=& \psi \, \chi \, (\pa_x \s ) \ . \label{eq:Bde2} \end{eqnarray} Recalling now that $\chi$ is, by assumption, a solution to \eqref{eq:Bde1}, these reduce to
\beq
\chi \, L(\psi) \ + \ \frac12 \ \[ 2 \, M(\chi ) \, M(\psi ) \ + \ \chi \, M^2 (\psi) \] \ = \ 0 \ , \ \ \ \
\chi \, M(\psi) \ = \ 0 \ . \label{eq:Bde3} \eeq
The second equation requires $\psi \in \mathrm{Ker} (M)$, and assuming this the first ones reduces to $\psi \in \mathrm{Ker} (L)$. Thus $\psi$ is any function in \beql{eq:Bker} \mathcal{F} \ = \ \mathrm{Ker} (L) \cap \mathrm{Ker} (M) \ . \eeq

Looking back at \emph{Example 7}, in that case $f = y p(t)$, $\s = y$; hence (with the present notation)
$ L = \pa_t  +  [p(t) - 1/2 ] x \pa_x$, $M  =  \pa_w  +  x \pa_x$.
It is then immediate to obtain that
$ \mathrm{Ker} (L) $ consists of arbitrary functions $P [z_\ell , w]$, while $\mathrm{Ker} (M)$ consists of arbitrary functions $Q [ z_m , t ]$ where
$$ z_\ell \ = \ x \, \exp \[ - \, \int_0^t [p(u) - 1/2 ] \, d u \] \ , \ \ \ z_m \ = \ x \, e^{-w}  \ . $$
In this case,
$\mathcal{F} = \mathrm{Ker} (L) \cap \mathrm{Ker} (M)$ is non-trivial, and consists of arbitrary functions \beq \eta \( x \ \exp \[ - \, w \ - \ \int_0^t [p (u)  - 1/2 ] \, d u \] \) \ . \eeq
To be more concrete, e.g. with the choice $p(t) = 1$ we get $\mathcal{F} =  \eta \( x \ e^{- w - t/2} \)$, while for $p(t) = t$ we get
$\mathcal{F} =  \eta \( x \ e^{- w  + t/2 - t^2/2 } \)$.

Let us now look at \emph{Example 8}. In this case $f=1$, $\s = y$; hence (with the present notation)
$ L = \pa_t +  (1 - x/2) \pa_x$, $M  = \pa_w + x \pa_x$.
In this case we get that again
$ \mathrm{Ker} (L)$ and $\mathrm{Ker} (M)$ consist respectively of arbitrary functions $P [z_\ell , w]$ and $Q [ z_m , t ]$, where now
$$ z_\ell \ = \ (x- 2) \, e^{t/2} \ , \ \ \ z_m \ = \ x \, e^{-w}  \ . $$
With these,
$ \mathcal{F} $ is trivial, i.e. reduces to constant functions.

\end{appendix}

%\newpage

\end{document}